%% file: 0.main.tex
\documentclass[acmsmall]{acmart} 

\usepackage{subcaption}

\AtBeginDocument{%
  \providecommand\BibTeX{{%
    \normalfont B\kern-0.5em{\scshape i\kern-0.25em b}\kern-0.8em\TeX}}}

\setcopyright{acmcopyright}
\copyrightyear{2018}
\acmYear{2018}
\acmDOI{10.1145/1122445.1122456}

\usepackage{booktabs} 
\usepackage{url}

\usepackage{caption}
\usepackage[all]{nowidow}
\usepackage{wrapfig}
\usepackage{array}
\usepackage{arydshln}
\usepackage{tabularx}
\usepackage{multirow}
\usepackage{arydshln}
\newcolumntype{L}[1]{>{\raggedright\let\newline\\\arraybackslash\hspace{0pt}}m{#1}}
\newcolumntype{C}[1]{>{\centering\let\newline\\\arraybackslash\hspace{0pt}}m{#1}}
\newcolumntype{R}[1]{>{\raggedleft\let\newline\\\arraybackslash\hspace{0pt}}m{#1}}

\usepackage{wrapfig,lipsum,booktabs} 

\usepackage{tabularx} 

\def\authnotes{1}
\newcounter{notectr}[section]
\newcommand{\thenote}{\thesubsection.\arabic{notectr}\refstepcounter{notectr}}

\newcommand{\note}[2]{$\ll$#1~\thenote: #2$\gg$}
\newcommand{\cnote}[1]{\ifnum\authnotes=1 \textcolor{blue}{\note{Comment:}{#1}}\fi}

\acmYear{2024}
\copyrightyear{2024}
\acmConference[26]{}{October 9--11, 2026}{USA}
\acmBooktitle{December 9--11, 2026, USA}
\acmDOI{10.1145/3700794.370XXXX}
\acmISBN{979-8-4007-XXXX-4/24/12}

\begin{document}



\title[DAIEM]{`\textit{DAIEM'}: Decolonizing Algorithm's Role as a Team-member in Informal E-market}

\author{ATM Mizanur Rahman}
\affiliation{
  \institution{University of Illinois Urbana-Champaign}
  \city{Urbana}
  \state{Illinois}
  \country{USA}}
\email{amr12@illinois.edu}

\author{Md Romael Haque}
\affiliation{
  \institution{Purdue University Fort Wayne}
  \city{Fort Wayne}
  \state{Indiana}
  \country{USA}}
\email{haque28@purdue.edu}

\author{Sharifa Sultana}
\affiliation{
  \institution{University of Illinois Urbana-Champaign}
  \city{Urbana}
  \state{Illinois}
  \country{USA}}
\email{sharifas@illinois.edu}

\renewcommand{\shortauthors}{Rahman et al.}

\begin{abstract}
In Bangladesh’s rapidly expanding informal e-market, small-scale sellers use social media platforms like Facebook to run businesses outside formal infrastructures. These sellers rely heavily on platform algorithms—not just for visibility, but as active collaborators in business operations. Drawing on 37 in-depth interviews with sellers, buyers, and stakeholders, this paper examines how people in informal e-market perceive and interact with the algorithm as a ``team member" that performs sales, marketing, and customer engagement tasks. We found that while sellers and local tech entrepreneurs are intrigued to develop services to support this industry, buyers and investors of the industry put their greater trust in human interactions. This surfaces a postcolonial tension associated with cultural values, local tech education and training, and a mismatch between the global and Bangladeshi e-markets' growth. We expand this discussion from multiple ongoing HCI, political design, and AI design angles. We also address the postcolonial tension and support the decoloniality movement in informal e-markets by proposing the \textit{DAIEM} framework that consists of six components: autonomy and agency; resistance; locality, culture, and history; rationality; materiality; and advocacy. Supporting the decoloniality and informality sentiment in informal e-market and other similar sectors, \textit{DAIEM} will serve both as a guideline for algorithm design and as an analytical tool. 

\end{abstract}







\settopmatter{printfolios=true}

\maketitle

\input{1.intro}

\input{2.lit}
\input{3.methods}

\input{4.context}

\input{5.find} 
\input{6.learnsales} 
\input{7.discussion}



\bibliographystyle{ACM-Reference-Format}
\bibliography{sample-base}

\end{document}

%% file: 1.intro.tex
\section{Introduction}
Across the globe, informal e-markets are flourishing on social media platforms like Facebook, Instagram, and WhatsApp \cite{emarket-1, emarket-2, emarket-3}. These platforms provide sellers with low-barrier tools to start and sustain small-scale businesses, particularly those excluded from formal market structures \cite{emarket-4, emarket-5}. From home-based clothing retailers in Nigeria to artisanal food vendors in the Philippines, sellers are leveraging familiar social media affordances to post products, chat with customers, and arrange deliveries \cite{nigerian, philipino}. In these settings, platform algorithms play a critical but often opaque role in determining posts and pages' visibility, traction gaining, and communication between the buyers and sellers. Thus, algorithms' role surpasses beyond being passive infrastructures and shape the interactions, visibility, and trust practices of sellers navigating informal digital economies.

This dynamic is particularly pronounced in Bangladesh's informal e-market, where platforms like Facebook, Whatsapp, and Telegram serve as the primary venues for a growing number of small entrepreneurs, especially women, youth, and first-time sellers \cite{bd-emarket, bd-women, bd-women2}. Here, sellers strategically engage with platform algorithms not just as invisible forces but as collaborators, designing their posts, responses, and promotional tactics around how algorithms boost or suppress visibility. Yet, this collaboration is double-edged: while algorithms help sellers reach audiences and grow their business, they also impose platform-specific pressures and introduce unpredictable breakdowns in visibility and communication \cite{bhuiyan2024curious, hoque2020customer}. Noteworthy that just not the sellers, but buyers and other stakeholders also engage with the algorithms of these social platforms and algorithms of other information and communication technologies (ICTs) associated with this informal e-market (i.e., \textit{Pathao Parcel} used by Bangladeshi gig-delivery workers \cite{pathao_parcel}).  

This research is motivated by HCI and CSCW scholarship that examines informal economic participation in the Global South through lenses of trust, mobile access, and vernacular entrepreneurship \cite{chandra2019rumors, rohanifar2022kabootar, sambasivan2012understanding}. This literature has explored how informal sellers leverage community norms, kinship ties, and mobile tools to manage trade, and how gender and identity shape access to platform infrastructures. Human-centered AI scholarship has also raised concerns about algorithmic opacity and labor exploitation in formal gig work \cite{weber2023opaque}. However, much of this work focuses on either formalized gig platforms or high-resource AI environments. Few studies have examined how informal sellers engage with algorithms not as extractive tools, but as strategic collaborators within resource-constrained, semi-automated ecosystems like Facebook commerce groups \cite{bhuiyan2024curious, hoque2020customer, hride2022linking}. Note that Facebook has lately shifted its focus toward being more market- and business-oriented over being just a social interaction platform, and informal e-market people in Bangladesh and many other countries in the Global South leaverage this platform's algorithmic priorities. However, how the algorithms of Facebook and other associated applications might be most effective in supporting the plaforms' and informal e-market's agenda is still understudied in HCI and social computing.

This paper addresses that gap by conceptualizing the algorithm not simply as infrastructure, but as an active participant in the inform e-market ecologies. Through an interview study (n=37) with informal e-market sellers, buyers, and related stakeholders in the Bangladeshi social media market we investigated their perspectives, practices, and tensions. While researching how algorithmic systems are interpreted, adapted, and sometimes resisted in everyday business operations, we addressed the following research questions: 
\begin{quote}
\textit{RQ1:} How do sellers and buyers in Bangladesh's informal e-market conceptualize and interact with algorithms of ICTs? \\ 
\textit{RQ2:} In what ways do they treat the algorithm as a strategic collaborator or teammate, and what tactics emerge from this perception? \\
\textit{RQ3:} What tensions, frictions, and emotional consequences arise from relying on algorithmic systems in many different ICTs in informal e-market? \\
\textit{RQ4:} How can the algorithms in different ICTs in informal e-market be more market efficient and supportive of this and similar other informal industry's growth?
\end{quote}

We found that sellers with limited access to formal supply chains and digital marketing staff are adapting to algorithmic logics of the platforms, bots, and other applications associated with their business, as if the algorithm itself were a ``team member" that amplifies posts, coordinates outreach, and facilitates real-time interactions with buyers. On the other hand, buyers put their trust more into human interactions over bots. Also, we noted both buyers' and sellers' urge for practicing their perceived agency, rationality, and autonomy and voice for adapting and continuing the existing cultural and materiality practice of the market. Additionally, the investors and entrepreneurs reported a mismatch of visions that raised concerns regarding this market's sustainable growth. This conversation and analysis further extend toward a postcolonial challenge that this ecosystem is currently facing. Building on this set of findings, we developed the decolonial algorithm for informal e-market (DAIEM) framework that would work as both a critical guideline for designing informal e-market-oriented infrastructure and an analytical tool for evaluating such ecosystems. 

Our work contributes to CSCW, HCI, and human-centered AI in four ways. First, we offer an empirical account grounded in rich interviews with sellers, buyers, and stakeholders in the Bangladeshi informal e-market and portray their voices, practices, and negotiations with algorithms in different ICTs associated with this ecosystem. Second, we introduce the concept of the algorithm as a ``sales team member,'' broadening human-centered AI scholarship to include informal, low-resource contexts of algorithmic labor. Third, our further analysis mapped the challenges algorithms create for different parties and how these parties find decolonial workarounds. Finally, we contribute by extending this discussion to develop a framework for designing decolonial algorithms (DAIEM) to support this informal e-market and similar other informal ecosystems.

%% file: 2.lit.tex
\section{Related Work}
The International Labour Organization (ILO) defines informal market as: all economic activities by workers and economic units that are, in law or in practice, not covered or insufficiently covered by formal arrangements \cite{ilo}. Therefore, the activities operate outside the formal reach of the law, or the law is not applied or enforced, or compliance with the law is discouraged. The history of market started off informal. In early human societies, direct exchange, like bartering in local marketplaces, was fundamental before formal economic systems \cite{barter}. Pre-industrial informal markets were central to community life, operating outside formal rules. 

The Industrial Revolution (18th-19th centuries) brought structured economies, but informal activities persisted. The "informal sector" emerged in development economics in the mid-20th century. In the early 1970s, Keith Hart's Ghana research highlighted unrecorded urban work, coining "informal sector" \cite{hart1973informal}.  The ILO further defined this in the 1970s as employment outside formal contracts without protection, initially seen as temporary \cite{ilo2}. However, the informal economy persisted and grew globally, showing resilience. Late 20th-century economic crises and structural adjustment policies under global development increased informalization as people sought income, underscoring its safety net role \cite{chen2007rethinking}. Globalization and outsourcing also fueled informal work and enabled new informal work via platforms, especially in developing nations \cite{oecd}. In today's world, examples of prominent informal market activities include independent ride-sharing, contractors in skilled trades working without formal company affiliation, direct sales of goods and services through informal channels, small-scale agriculture and local food processing, and street vending \cite{husain2015assessment, wiego}.

The \textit{informal e-market} is a relatively new concept in both business and also in scholarship. This can be defined as "online economic activities characterized by transactions that occur largely outside formal regulations, often leveraging digital platforms like social media, messaging apps, or basic online listings rather than dedicated e-commerce websites with integrated payment and delivery systems" \cite{inf-ecom}. Informal e-markets evolved from early remote transactions via teleshopping and the nascent World Wide Web, which lacked structured regulations. The rise of social platforms saw users engaging in direct, person-to-person sales, bypassing formal e-commerce frameworks. Mobile internet growth amplified this "social commerce" on platforms like \textit{Instagram} and \textit{WhatsApp}, where deals were often based on personal trust and basic payment methods, outside established online marketplaces \cite{emarket-1, emarket-03}. The thriving informal e-markets of the 2020s, often on social media marketplaces and community groups, continue this trend with direct negotiation and decentralized logistics, operating with minimal formal oversight compared to regulated e-commerce platforms \cite{undp, naghavi2019social}. While informal e-market is traditionally relationship-driven, it is increasingly influenced by data-driven technologies and AI tools to manage targeted advertising within social media groups and initial customer interaction\cite{wu2025ai, aljarboa2024factors}. Below, we first provide an overview of the informal e-market in broad terms and then focus on the Global South scenarios. Later, we look into the intersection of algorithm, collaboration, and informal e-market.  

\subsection{Global Informal E-market}
Some of the most popular informal e-markets are social media-based. For example, \textit{Facebook Marketplace} and Groups stand out globally for facilitating local buying and selling, often involving individual and small-scale informal sellers \cite{fb1,fb2}. Its user-friendly interface and extensive network make it a significant hub for these transactions. \textit{Instagram} serves as a visual storefront for informal commerce, particularly in fashion, crafts, and food, with sales often conducted through direct messaging and less formal payment methods \cite{insta1, insta2}. Messaging apps like \textit{WhatsApp} and \textit{Telegram} act as direct sales channels, enabling small businesses and individuals to connect with customers, share product details, and arrange transactions on a personal level \cite{tandw1, tandw2, tandw3}. \textit{TikTok}'s growing influence in "social commerce" includes informal sales through live streams and direct interactions, appealing strongly to younger demographics \cite{tiktok1, tiktok2}.

North American community platforms like \textit{Craigslist}, \textit{Nextdoor}, \textit{OfferUp},  \textit{eBay Classifieds}, \textit{VarageSale}, and \textit{5miles} facilitate diverse local transactions, offering a wide range of classified advertisements for buying, selling, and services, often involving direct interactions \cite{lit1-minkus2014know, lit2-duh2002control, lit3-choksi2024under, lit4-masden2014tensions, lit7-kurwa2019building}. In China, \textit{Taobao} is a major e-commerce platform that hosts numerous smaller individual sellers who may operate with varying degrees of formality \cite{china_tab}. Southeast Asia's \textit{Shopee} and \textit{Lazada}, while established e-commerce players, also facilitate transactions for many small, potentially informal sellers, especially as they grow \cite{shop_and_laz, shop_and_laz2}. \textit{Mercado Libre} is a prominent platform in Latin America, similar to \textit{eBay} and \textit{Amazon}, hosting a diverse array of sellers, including numerous small and informal businesses \cite{mercado}. \textit{Jumia} in Africa serves as a pan-African e-commerce platform that enables informal traders to reach a broader customer base, sometimes bridging the gap between informal and formal commerce with support for logistics and digital payments \cite{jumia}. A vast literature in HCI and social computing addresses how advertisement, post boosting, and paid content on different social media influence users' interaction on respected platforms \cite{voorveld2018engagement, shahbaznezhad2021role, eslami2016first}. We advance this literature by bringing in a multi-dimensional perspective of sellers, buyers, investors, delivery personnel, and other stakeholders and depict how algorithmic raises postcolonial concerns (RQ1, RQ2, RQ3).

\subsection{Global South, Informal E-Markets, and Vernacular Entrepreneurship}
Informal e-markets in the Global South showcase vernacular entrepreneurship, where digital technologies are adapted to local norms, infrastructures, and labor practices. Research on South Asian marketplaces \cite{chandra2017bazaar, pal2018digital} highlights how sellers build trust and coordinate operations through embodied practices like bargaining and product testing. These markets often resist platformization, creatively using tools such as intercoms or WhatsApp to maintain informal coordination and community control, rather than adopting standardized e-commerce infrastructures. This resistance is a form of translation, as seen with textile sellers in Surat, India, adapting rigid e-market workflows into flexible WhatsApp exchanges, enabling dynamic, women-led collaborations in cataloging, order processing, and distribution based on local social norms \cite{joshi2025reselling}. Similarly, Nairobi youth on Facebook consolidate diverse digital practices for job searches, remittances, and service promotion, driven by cost, data, and infrastructural constraints \cite{wyche2013hustling}. The Global South literature often portrays digital technologies as uplifting informal workers, but Bhattacharya \cite{bhattacharya2019ict} cautions against universalizing assumptions, noting that informality can be a rational response to structural exclusion, with actors prioritizing stability and privacy over scale. Participatory and context-aware designs, like the Machinga app for Tanzanian street traders \cite{rumanyika2022design}, and addressing infrastructural precarity \cite{ebrahim2024barriers} are crucial for effective support. Gender also plays a significant role, with platform formalization and mobile money solutions sometimes failing to align with the realities of women entrepreneurs \cite{de2025whatsapp, mustafa2019digital}, highlighting infrastructural exclusion. Transnational communities also develop informal digital solutions to overcome financial exclusions \cite{rohanifar2021money, rohanifar2022kabootar}.

In the Bangladeshi context, sellers on social media, while benefiting from interactional intimacy, face trust and logistical challenges. Consequently, many supplement their social media presence with websites to project professionalism and credibility. This dual-platform strategy demonstrates a layered approach to digital entrepreneurship in informal markets, prioritizing strategic tool use over single-platform streamlining. Studies indicate that while social media offers visibility, Bangladeshi (and also Nigerian) consumers tend to trust websites more for transactions \cite{clemes2014empirical, yahaya2021mediating}, with perceived professionalism being locally defined, as seen in Thai online grocery markets where personalization builds buyer trust \cite{driediger2019online}. Research in Dhaka confirms that visible trust cues, such as social validation and engagement, significantly influence consumers, often more than purely functional comparisons \cite{enam2024online, suhan2015acceptance}. Even non-commerce focused studies in Bangladesh highlight the deep integration of ICTs in communication and reputation building, especially in environments with limited institutional support \cite{hia2020use}. This set of literature motivates us to investigate how Bangladeshi informal e-market people employ their situated knowledge, social connections, and custom online activities to overcome the challenges caused by standardized e-market norms (RQ1, RQ3).

\subsection{Algorithmic Collaboration in Informal Labor}
CSCW research on platform labor has critically examined algorithmic management in gig work, highlighting the social, emotional, and tactical labor involved in navigating opaque systems. Sehrawat et al. conceptualize ``professional durée" to describe how Uber drivers in Hyderabad gradually attune to platform rhythms \cite{sehrawat2021uber}. Weber et al. expand on this with a typology of sensemaking strategies, such as experimenting, gossiping, and resisting. Yet little work explores algorithmic interaction in informal, non-gig commerce \cite{weber2023apathy, weber2023opaque}. Chandra and Pal show how vendors in Bangalore use rumors to interpret platform disruptions \cite{chandra2019rumors}. Pal et al. observe how demonetization initiatives clashed with embedded trust networks, causing vendors to temporarily adopt and then abandon digital payment solutions \cite{pal2018digital}. Research has also revealed how sellers perceive and adapt to Facebook's algorithmic dynamics. Tactics such as post-timing, engagement optimization, and ad boosts are treated as “collaborative” interactions with the algorithm. Buyers, in turn, assess legitimacy based on comment counts and follower numbers—patterns that align with Sambasivan et al.'s work that emphasize negotiation and informal accountability in shared ICT use \cite{sambasivan2012understanding}. Mathur et al. further highlight how affiliate marketing obscures commercial intention on platforms like YouTube, raising broader questions about algorithmic visibility and disclosure \cite{mathur2018endorsements}. Hossain showed that mobile-first platforms double as branding infrastructures in emerging markets, entangling algorithmic collaboration with self-presentation and entrepreneurial labor \cite{hossain2019social}. This body of literature foreground our understanding of postcolonial algorithmic influence in Bangladeshi informal e-market (RQ3) and motivate us to develop the DAIEM framework to support the sustainable growth of informal e-market in Bangladesh and similar other settings (RQ4). 


%% file: 3.methods.tex
\section{Methods}
To solicit the answers to our four research questions, we conducted in-depth interviews with 37 participant. All participants and researchers in this project were born and raised in Bangladesh and are native Bengali speakers. This shared background allowed for richer communication, cultural sensitivity, and deeper contextual understanding throughout the data collection and analysis process. The study was reviewed and approved by the Institutional Review Board (IRB) at the authors' institution. Below, we describe our method and the process of data collection and analysis.

\input{3.tab-demo}

\subsection{Participant Recruitment}
We leveraged our social capital to identify initial participants from both seller and buyer communities active in informal e-market. To broaden our reach, we also circulated a public post on Facebook explaining the study and inviting interested individuals to contact us directly. From there, we proceeded through snowball sampling, where participants referred others who might be a good fit for the study \cite{goodman1961snowball}. To be eligible, participants needed to meet three selection criteria: (a) be 18 years or older; (b) reside in Bangladesh; and (c) have prior experience with informal e-market platforms (e.g., Facebook, WhatsApp, Instagram, etc.) either as a seller, buyer, or relevant stakeholder. These criteria were mentioned in the recruitment flyer and also discussed with interested individuals during initial contact. The pool of participants included sellers and buyers of informal e-market, four entrepreneurs who develop AI tools for informal e-market, two investors in startups that develop low-cost support-bots for such informal businesses, and one admin of a support group that serves informal e-businesses. See Table~\ref{Tab:demo} for summarized details of the participants (and supplementary materials for more). Interested participants contacted us after seeing the flyers. In cases of snowball sampling, we also contacted some participants, such as some sellers, a support group admin, and investors.

\subsection{Semi-Structured Interviews}
All our semi-structured interviews (n=37) took place upon seeing participants' consent \cite{ruslin2022semi}. Each interview lasted between 30 to 60 minutes and was scheduled based on the participant's convenience. Interviews were held remotely over Zoom, as the first author was based in North America during data collection. The interviews were conducted in Bengali, the native language of all participants and researchers. In some cases, we used a translated version of the interview protocol to ensure clarity and comfort during the conversation. At the beginning of each session, we explained the purpose of the study, emphasized the voluntary nature of participation, and obtained oral consent. Participants were informed about the ethical considerations, including assurances that their identities would remain anonymous and that they could withdraw at any point without penalty.

Our interview questions focused on understanding participants' interactions with algorithms of the platforms and other applications used in informal e-market (e.g., Whatsapp, Pathao Parcel, Odol-bodol etc.). We asked the sellers about their strategies for managing visibility and customer engagement, and the challenges they encounter in informal e-market ecosystems. This included discussions on algorithmic tactics, logistical constraints, situated knowledge, and local social and trust dynamics that influence their business practices. Our interview with buyers' focused more on their interaction and challenges with different human and algorithms in this ecosystem and their workarounds. The interviews with the investors and admin of support group for informal e-marketers emphasized on mission and visions of the market and the roadblocks to this ecosystem's sustainable growth. All sessions were either audio-recorded with participant consent or documented through detailed note-taking. These recordings and notes were later transcribed and used for open coding and thematic analysis \cite{fereday2006demonstrating, boyatzis1998transforming}.

\subsection{Data Collection and Analysis}
The interviews generated approximately 21 hours of audio recordings and 223 pages of transcribed and translated data. After each session, audio recordings were transcribed and translated from Bengali to English. All identifying information was removed during the transcription process to maintain participant anonymity. We followed an open coding and thematic analysis approach to analyze the interview data \cite{fereday2006demonstrating, boyatzis1998transforming}. Two members of the research team independently reviewed the transcripts to familiarize themselves with the data. Through an open coding process, we allowed codes to emerge inductively, capturing participants’ experiences, strategies, and concerns related to informal e-market practices. The initial codes were platform flexibility, visibility advantage, profit retention, brand identity, among others. The researchers then discussed the codes in several collaborative sessions, identifying patterns and grouping related codes into broader themes. This iterative process allowed us to refine our codebook (see supplementary material) and develop a set of high-level themes that represent key aspects of participant experiences. These themes form the basis of our findings and are presented in detail in the following section.




%% file: 3.tab-demo.tex
\begin{table}[!t]
\begin{center}
\begin{tabular}{|rl|}
\hline
Total Participants: & 37 (Female: 20, Male: 17)\\
\hdashline
\multicolumn{2}{|c|}{\textbf{Participant Roles}} \\
\hdashline
Buyer (only): & 10 (Female: 6, Male: 4)\\
Seller (only): & 12 (Female: 9, Male: 3)\\
Buyers \& Sellers (both): & 6 (Female: 3, Male: 3)\\
Sales Bot Entrepreneur: & 4 (Female: 1, Male: 3)\\
Investor: & 2 (Female: 0, Male: 2)\\
Online Support Group Admin: & 1 (Female: 1, Male: 0)\\
Delivery Personnel: & 2 (Female: 0, Male: 2)\\
\hdashline

\multicolumn{2}{|c|}{\textbf{Age Range (in Years)}} \\
\hdashline
All: & 18-56, median 26\\
Female: & 19-55, median 26\\
Male: & 18-56, median 25\\
\hdashline

\multicolumn{2}{|c|}{\textbf{Education}} \\
\hdashline
Higher Secondary School: & 7 (Female: 3, Male: 4)\\
Pursuing Bachelor's Degree: & 10 (Female: 3, Male: 7)\\
Bachelor's Degree: & 15 (Female: 12, Male: 3)\\
Pursuing Master's Degree: & 2 (Female: 1, Male: 1)\\
Master's Degree: & 2 (Female: 0, Male: 2)\\
Pursuing PhD: & 1 (Female: 1, Male: 0)\\
\hline

\end{tabular}
\end{center}
\caption{Details of the Participants}
\vspace{-15pt}
\label{Tab:demo}
\end{table}

%% file: 4.context.tex
\section{Context: Introducing Informal E-market in Bangladesh}
Bangladeshi informal e-market is dominated by small-scale entrepreneurs who leverage the local networks on social platforms such as \textit{Facebook}, \textit{Instagram}, and \textit{WhatsApp}. These platforms allow anyone to start selling instantly without formal registration or tax obligations, providing a quick and flexible entry point for small businesses and young entrepreneurs. Formal e-markets ensure accountability through structured logistics, digital tracking, and standardized dispute resolution, while informal e-markets rely on trust-based communication and ad hoc delivery networks that offer flexibility but limited traceability. Informal sellers retain full pricing control, enabling quick adjustments and personalized discounts without formal deductions. Informal e-markets foster more personalized engagement via live videos, social posts, and direct messaging. 

\subsection{Advantages in Informal E-Market}
Informal e-markets offer a range of benefits for both sellers and buyers, creating a flexible and personalized marketplace. This subsection details the main advantages that attracted our participants.

\subsubsection{Relaxed Verification Policy, No Registration Cost, and No Profit Sharing} 
The key factors for choosing \textit{Facebook}- and \textit{Whatsapp}-base market are that it is easy to start a business in informal e-market, the verification and registration processes are almost zero work and does not require paperwork if the business uses Facebook Page, Facebook Group, or\textit{ Whatsapp} Group options and does not link any online payment mechanism. Also, the reach of these platforms are widespread use across the country. Even people in rural areas or older age groups, use \textit{Facebook} daily for personal and social purposes. This familiarity helps sellers reach their target audience, as P12 explained:

\begin{quote}
\textit{``Starting a business on Facebook is easier because this is the social media with the largest number of users. Not all people use Daraz or Bikroy.com, but I think in our country everyone who has access to the internet does use Facebook a lot. So, it is a big market to capture and start your business.", \textbf{(P12)} }
\end{quote}

Another reason that the sellers relied on this informal e-market was that they did not need to share their profits with the platforms, which would have been a case if they enrolled their business on a formal platform. On formal platforms, a commission is taken from each sale, reducing the seller's profit. For those operating on small budgets and narrow margins, these fees are often unsustainable. Seven participants shared that the commission deductions on formal platforms create a heavy burden. In contrast, on \textit{Facebook}, sellers keep the entire revenue, which is a significant advantage. 

\subsubsection{Autonomy and Identity} 
All the sellers told  us that they enjoyed the \textbf{autonomy} of managing their business using \textit{Facebook} option. The pages allowed customizing the layout and choosing the language (Bangla or English) to deciding how they interact with customers and set the tone for their business. This level of independence fostered creativity and allowed them to be more responsive to customer needs. Participant P5 shared how this freedom influenced her decision, noting that having complete control over her page made her feel more confident and independent in running her business.

\begin{quote}
\textit{``I have been familiar with Facebook for years. Creating a page and customizing on your own accord seemed easier here. On formal platforms, their policies and communication felt restrictive. Here, I can update my page whenever and however I want.", \textbf{(P5)}}
\end{quote}

This financial independence in informal e-market also links to another motivation: \textbf{establishing own identity}. Eight participants considered their business as a part of their identity. They wanted customers to recognize their name, their brand, and their unique offerings. Formal platforms make it harder to establish this recognition, as the platform name often overshadows the seller's identity. P20 shared her thoughts on how selling through \textit{Daraz} could affect their brand:

\begin{quote}
\textit{``For building my identity, I have long-term plans for this business. If I sell through Daraz, people will think they bought the product from Daraz, not from me. My name or brand will not be recognized. I want people to remember me when they use my product and to know who I am and recognize my brand.", (\textbf{P20})}
\end{quote}

The benefits of autonomy and self-promotion offered by \textit{Facebook} and \textit{Whatsapp} groups made informal e-market ideal platform for sellers looking to build a distinct identity. Unlike larger, more rigid market platforms, \textit{Facebook} allows direct connections with customers and full control over how products are presented and marketed, which was advantageous to the sellers. 

\subsubsection{Real-time Phone Calls Leading to Trust, Reliability, and Connections} 
Nine buyers said they prefer informal e-markets because they allow real-time chats and calls with sellers, verify product details, and request customizations, which boosts their confidence in making purchasing decisions. The real-time conversations make it easy for buyers to compare sellers and products on \textit{Facebook}. In groups and pages, multiple sellers showcase similar items, allowing buyers to browse, ask questions, and gather information quickly. This direct communication lets them negotiate prices and judge a seller's \textbf{reliability} based on responsiveness. 

Our participants also discussed \textbf{trustworthiness} in purchases on online informal market. Twelve buyers shared that once they find a seller they trust, they prefer returning for future purchases. This ongoing \textbf{connections} helps buyers and sellers understand each other's preferences, communication style, and expectations, making transactions smoother and more confident. As Participant P10 noted, positive experiences with a seller encourage repeat business.

\begin{quote}
\textit{``Typically I trust buying from people whom I already know. There is no trust issue and it is kind of sure that I will get a good product and even if I do not, at least I have an option to ask for a product exchange." \textbf{(P10)}}
\end{quote}

This focus on personal connections and trust makes informal platforms appealing to buyers who value long-term relationships with sellers. The ability to establish bonds, resolve issues directly, and rely on familiar sellers offers a significant advantage over more impersonal formal platforms.

\subsection{Challenges in Informal E-Market}
This subsection highlights some common drawbacks that can hinder transactions, restrict growth, and undermine trust in this rapidly growing marketplace.

\subsubsection{Intense Competition and Distrust} 
Sellers told us that they face intense competition on social media platforms. The simplicity of starting a business page leads to an influx of new sellers daily, making it difficult to stand out. Larger, established sellers can post frequently, create high-quality visuals, and pay for advertising, attracting more followers and engagement. In contrast, smaller sellers, managing everything on their own, often struggle to compete. This imbalance can create stress and a sense of unfairness. Eight sellers also mentioned that, unless they have an \textbf{established identity} or \textbf{loyal customer base}, their page struggles to gain visibility. Despite putting in significant effort, they find it difficult to reach a wider audience without relying on paid promotions. With the \textit{Facebook} marketplace becoming increasingly saturated, standing out in such a crowded space feels nearly impossible without continuous effort. To stay competitive, some sellers focus on frequently introducing new or creative products, believing that offering something unique is key to staying relevant. However, this strategy \textbf{demands unfairly constant innovation}, planning, and experimentation, which can quickly become exhausting.

In addition to competition, sellers face a persistent \textbf{lack of buyer trust}. Many Bangladeshi customers still prefer in-person shopping, viewing platforms like \textit{Facebook} as risky due to unclear standards for quality, returns, and accountability. Even isolated scams or accidental poor product experiences cause \textbf{mass hesitation}, shaped by cultural habits and skepticism toward unfamiliar pages. Trust is often tied to recognizable brands or strong social media presence, which takes time to build, especially in the absence of formal verification or platform support. As P28 explained:

\begin{quote} \textit{``In our country most people still prefer traditional in-person shopping. When someone opens a new page, buyers often assume it will not be good. They trust old, established pages with strong brand value.", \textbf{(P28)} } \end{quote}

Sellers noted that without formal verification and with the risk of scams, new businesses struggle to build credibility and gain customer confidence.

\subsubsection{Limited Search Functionality} 
Another challenge buyers face in the informal e-market is from their fragmented search experience. Search results are influenced by unpredictable algorithms, forcing buyers to browse through unrelated content, such as irrelevant profiles, personal posts, or random pages. Participant P29 shared how these limitations impact their shopping experience.

\begin{quote}
\textit{``If I search for an item on Facebook, what usually comes up are Facebook profiles, random pages, or unrelated videos. Sometimes, these results are misleading or even scams. But on formal platforms, when I search for an item, I get a complete list of matching results, which is very helpful." \textbf{(P29)}}
\end{quote}

Many buyers browse business group and trusted pages, or waiting for recommendations in their feed as workarounds. However, these approaches are slow and unpredictable.

%% file: 5.find.tex
\section{Findings: Algorithm as a Team Member in Informal E-market}
This section presents our findings and analysis regarding sellers', buyers', and other stakeholders' perspectives regarding current and possible future roles of algorithms in different ICTs associated with the informal e-market. Then we discuss the affordances that the informal e-market still expects from algorithms. 

\subsection{Sellers' Perspectives}
The algorithms in different ICTs associated with informal e-markets help sellers reach customers and manage marketing. However, they also struggle with tasks that require human judgment, emotional understanding, and personalized responses. This section discusses the sellers' perspectives on both the strengths and weaknesses of algorithms.

\subsubsection{Strengths of Algorithms as a Seller Team Member}
The algorithm offers several advantages to sellers by automating repetitive tasks, amplifying reach, and supporting customer engagement. In this subsection, we highlight some of the key strengths that make algorithms valuable team members in the informal e-market.

\textbf{a) Boosting Posts from Periodic Replies} In Bangladesh's informal e-market, many sellers intentionally withhold prices in their posts, while a human staff member might manually try to attract attention through marketing calls or customer reminders. The algorithm performs this task more efficiently by automatically boosting content that receives comments, messages, or reactions. Eleven sellers said that this way, they receive more comments, direct messages, and more engagement. Here, the algorithm serves as a conversation facilitator. When customers message the seller, it creates an opportunity for personal interaction, such as offering discounts, discussing preferences, or promoting new designs. P11 described how this leads to longer-term benefits:

\begin{quote}
\textit{``When a customer messages me to ask the price or comments, that boosts engagement. When someone messages me directly, I sometimes offer a discount. For example, if a product is 300 taka, I might offer it for 230 or 250. They are happy, and if I understand their preferences, I can later show them new products or designs. Often they end up buying from me again.", \textbf{(P11)}}
\end{quote}


The seller participants noted that the algorithm could automatically amplify content based on patterns of interaction. Nine sellers shared that the first few hours after a post are especially important, as quick engagement, like comments, likes, and shares, signals interest to the algorithm, leading to more visibility. Sellers who see the algorithm as a teammate carefully plan their replies and self-comments to stretch this attention window, extending the life of the post across more newsfeeds. P7 explained how she carefully times her responses to maximize platform support.

\begin{quote}
\textit{``When someone comments asking for the price, I wait about ten minutes before replying, then delete that comment and reply again after another thirty minutes. I also comment on my own post and ask my friends to comment. The goal is to keep the post active in the newsfeed, especially during the first two to three hours, to reach more buyers.", \textbf{(P7)} }
\end{quote}

Thus, sellers are aware that platform algorithms can monitor in real-time, adjust visibility based on response trends, and reward activity with expanded reach. They use the algorithm to trigger network-based reach, where a single like or comment can make the post visible to the entire friend network of that user. 



\textbf{b) Paid Boosting as Algorithmic Amplification} Thirteen sellers mentioned that they used paid boosting and saw that the algorithm helped them define their audience by age, location, gender, interests, and behaviors. Seven sellers also described how the algorithm rewards this collaboration by keeping boosted posts visible longer and enabling more meaningful engagement. Once a boosted post gains attention, it triggers further reactions, shares, and private messages, signaling to the algorithm that the post is valuable. This feedback loop continues to drive reach even after the initial promotion ends, creating ripple effects for hours or days. Participant P31 explained how a small budget helped her build lasting customer networks.

\begin{quote}
\textit{``Sometimes even boosting for just 100 or 200 taka helps me reach thousands. If the product is good and the audience is right, they message me, follow my page, or even share the post. Then, Facebook continues to show the post without more boosting.", \textbf{(P31)}}
\end{quote}

Compared to formal e-market systems with fixed listing and ranking rules, \textit{Facebook}'s boosting system allows sellers to test different strategies, track results, and optimize performance over time. Participant P14 emphasized how this algorithmic partnership gave her a better understanding of her audience and helped refine her approach.

\begin{quote}
\textit{``When I boost a post, I can see which types of people are clicking, who is messaging, and what kind of product is getting more attention. It is really beneficial for me because I get to understand what I should do better next time", \textbf{(P14)}}
\end{quote}


\textbf{c) Automated Chatbots as Sellers' First Responder} Several small-scale sellers mentioned that they manage everything on their own, and hence, being available 24/7 is nearly impossible. Ten sellers told us they used chatbots to respond instantly to common buyer queries about price, availability, or the ordering process. Unlike a human assistant who might need rest or can only handle a few conversations at a time, the chatbot works around the clock, handling multiple queries simultaneously, ensuring no potential buyer is ignored. P28 described how this helped prevent lost sales.


This ability to maintain real-time engagement, even during off-hours, makes the chatbot a crucial support system for sellers. Moreover, unlike formal platforms with rigid customer service structures, informal sellers on \textit{Facebook} can customize their chatbots with personalized greetings, product links, and a tone that reflects their brand. As P22 said:

\begin{quote}
\textit{``I set up the chatbot to reply the way I normally would. When someone asks about a product, they get a greeting, the price, and a link to order. Before, I had to type all that myself every time. Now, I can focus on packing and delivery while the chatbot handles the initial questions from buyers.", \textbf{(P22)}}
\end{quote}

In these ways, automated chatbots serve as front-line responders on behalf of sellers, ensuring that customer interest is captured, acknowledged, and nurtured when the seller is offline or busy. 


\textbf{d) Urgency as Strategy for Outreach} Several of the sellers mentioned that they often played the platform algorithm by using urgency cues like limited-time offers, free delivery, or seasonal discounts. These urgency signals act as triggers, prompting the algorithm to promote the post more frequently in user feeds, reaching a much larger audience. Upon requesting further explanations, sellers told us that by writing captions that highlight time-sensitive opportunities or limited stock, sellers create a narrative that prompts immediate buyer interest. The algorithm then extends this reach, turning one well-written post into a wide-reaching campaign. This is particularly valuable for small sellers who cannot afford repeated paid ads or full-time marketing support. P18 described how his discount posts gained more reach thanks to the algorithm's reinforcement.

\begin{quote}
\textit{``I feel that posts where I write 'limited stock' or 'offer ends today' get more visibility. Those kinds of posts seem to reach more people automatically. Even if I do not boost them every time, they spread a lot more compared to normal posts. So now I try to add these urgency lines in my captions.", \textbf{(P18)}}
\end{quote}

These examples show how sellers strategically compose their posts to align with what the algorithm amplifies, turning it into a high-performing team member that enhances marketing efficiency in Bangladesh's informal e-market.

\subsubsection{Shortcomings of Algorithm as a Seller Team Member}
The algorithms of different ICTs in informal e-market also have limitations. Many tasks that require human judgment, emotional intelligence, or personalized responses remain beyond the reach of current algorithms, as we describe them below.

\textbf{a) Algorithmic Limitations and Platform Glitches} In theory, the algorithm should support smooth communication and post engagement. But in practice, it often misses tasks that a reliable human assistant could handle, like ensuring messages are seen or customer questions in comments are answered. Thirteen sellers reported notification failures, stating that they often do not receive alerts of customers' comments on their posts. In the fast-paced environment of informal e-markets, missing these interactions can cost a sale. Since the algorithm does not always flag these interactions in time, sellers have to manually check each post repeatedly to avoid missing potential buyers. P3 explained how this issue creates misunderstandings that harmed her reputation.

\begin{quote}
\textit{"Sometimes people comment on a post, but I do not get the notification due to a Facebook glitch, so I miss replying. That creates a negative impression, and customers might think the page is unprofessional. But it is not always my fault, but a platform issue." \textbf{(P3)}}
\end{quote}

Another recurring problem is message delivery failure. During high traffic, the platform algorithm sometimes blocks or delays sellers' responses to buyers, especially in \textit{Facebook} groups. This is frustrating when the algorithm allows buyers to send messages but prevents sellers from replying. 


\textbf{b) Platform Pressure and Emotional Exhaustion} Beyond visibility concerns, many informal sellers experience a deep emotional toll from the constant demands of maintaining their online presence. Six sellers shared that they need to post photos, reels, stories, and go live regularly to keep their pages visible, and even a short break can cause their reach to drop, making it feel like the algorithm is working against them. While each task may seem small, doing them repeatedly adds up, creating ongoing pressure. P5 described this as exhausting,

\begin{quote} \textit{``Facebook has some weekly challenges. They give instructions like posting a certain number of reels, videos, and live sessions each week, not just photos. There is also a new thing now, like posting 10 or 15 stories every day. Not completing these goals fades the boosting effect, and page engagement drops. They even ask for 100 new likes within 15 days, and if I do not keep up, they show a red signal indicating my page is in a downward position.", \textbf{(P5)}} \end{quote}

Also, the algorithm does not recognize religious holidays, personal emergencies, or burnout, and simply expects sellers to stay consistent. Thus, algorithms lack empathy and the contextual judgment that a human teammate could provide. Sellers are left to shoulder the emotional labor of staying visible, often at the cost of rest, family time, and mental peace. In this way, the algorithm's inability to adapt to human rhythms becomes a significant shortcoming of its role as a team member, placing sellers under continuous and often unsustainable pressure.

\textbf{c) Manual Verification Through Customized Product Videos} While algorithms and chatbots can automate replies or share generic product information, they struggle when buyers ask for specific, customized product videos to confirm authenticity. Sellers shared that many buyers hesitate to place orders based on standard photos or captions, often requesting personalized videos that show product quality, size, color, or packaging. Different buyers may want different angles or features, forcing sellers to record multiple videos for the same item. Despite chatbots providing quick standard replies, they cannot anticipate or fulfill these unique video requests. Participant P29 explained the difficulty of managing these expectations manually.

\begin{quote}
\textit{``Buyers often ask for detailed videos to make sure the product is genuine. I have to make different videos for different customers depending on their requests. Some want to see the fabric up close, others want size comparisons or packaging. I have to do it all myself, which takes a lot of time. Sometimes, I do not even understand exactly what kind of video the customer wants." \textbf{(P29)}}
\end{quote}


\textbf{d) Manual Comment Moderation and Community Engagement} One major shortcoming of the algorithm as a seller team member in the informal e-market is its inability to manage harmful or malicious comments. In environments like \textit{Facebook}, where comment sections are crucial for building trust. Sellers often face individuals who post negative remarks out of jealousy, personal grudges, or competitive motives, accusing them of dishonesty or questioning product quality. Since the algorithm cannot distinguish between sincere buyer concerns and hostile interference, sellers must search through each post, identify harmful feedback, and delete it individually, which is time-consuming and emotionally exhausting. P12 shared that this was stressful, 

\begin{quote}
\textit{"I have to check for these comments because if they stay up, they can hurt my business. Finding them is hard because I do not always get notifications, and I have to go through each post to hide or delete them. This takes a lot of time and feels discouraging, but I have to do it to protect my page's reputation." \textbf{(P12)}}
\end{quote}




\textbf{e) Understanding Buyer Mentality Through Manual Interpretation} Another shortcoming of the algorithm as a seller team member is its inability to understand the emotional and contextual nuances of individual buyers. Particularly, chatbots lack the emotional intelligence needed to connect with diverse customers. In informal e-markets, the success of a sale often depends not just on the product, but on how well the seller can tune into a buyer's mindset, gauge the tone of the conversation, and recall past interactions. This includes remembering what the customer purchased before, any complaints or compliments they shared, and how responsive or cautious they were. This kind of adaptive emotional labor requires careful attention, intuition, and patience, which current algorithms simply cannot provide. P16 explained the manual labor in these tasks, 

\begin{quote}
\textit{``... (I)t is important to understand their mindset. When they are about to buy something, I try to figure out what they like, their emotional tone, and sometimes even check their past chat history to see what they bought before or might want now. If I have a good sense of these things, I can keep them happy and sell my products successfully. These days, many customers message me, and it is not easy to understand everyone individually, but I try my best." \textbf{(P16)}}
\end{quote}

Sellers also noted that sometimes they visit a customer's profile to check for clues about their age, gender, lifestyle, or interests. These cues help them tailor product suggestions, communication styles, and negotiation approaches. For instance, a younger buyer might prefer trendier product options, while an older one may focus more on durability and value. Understanding these nuances allows the seller to customize their sales pitch in ways that feel personal and effective. However, this kind of individualized attention requires time, judgment, and cultural insight, which current automated systems lack.

\textbf{f) Disrupting Business with Automated Content Moderation Algorithm} Three participants told us about their experiences with automatic content warnings from the platform's community policies and moderation, and how those disrupted their business. P10 told us that his dry fish business was hampered, and he went through a hassle, 

\begin{quote}
\textit{``I posted close-up pictures of `shutki' on Facebook. After some time, I received a warning that my post was flagged for ‘prohibited items’ and ‘potentially dangerous goods.’ I do not exactly remember the exact wording, but Facebook might have mistaken the dried fish as an unregulated food item. After that, I noticed a significant drop in the reach of my other posts for the next few days." \textbf{(P10)}}
\end{quote}

P5 also told us about how her posts about her clothing business were flagged and promotion was disrupted because of having local cultural and indigenous visual components. 

\begin{quote}
\textit{``Sometimes, when I post photos of saris with colorful designs, my posts get flagged unexpectedly. I included images of people wearing traditional, comfortable clothing, but out of nowhere, the post was flagged for "nudity or sexually suggestive content." It makes no sense. This has happened two or three times, without any clear reason." \textbf{(P5)}}
\end{quote}

Thus, our participants expressed their frustration around platform algorithms' insensitivity to \textbf{culture, heritage, and local historic value} and further harm to their business.

\subsection{Buyers' Perspectives}
Algorithms on different ICTs in informal e-market also influence what buyers see, how they discover products, and even how they make purchasing decisions. In this section, we explore both the benefits and challenges of algorithms from the buyers' perspective, highlighting the ways they enhance convenience while also introducing new limitations.

\subsubsection{Strength of Algorithm: Buyers' Perspectives}
Algorithms provide several advantages to buyers by simplifying product discovery, personalizing shopping experiences, and reducing the time spent searching for items. In this subsection, we highlight some of the key strengths that make algorithms valuable digital assistants for buyers in the informal e-market.

\textbf{a) Algorithmic Integration into Everyday Scrolling} One key advantage of the algorithm as a team member is its ability to seamlessly blend shopping into everyday scrolling. In Bangladesh's informal e-market, buyers often discover products through algorithmically suggested posts while browsing \textit{Facebook}, turning shopping into a passive, low-effort activity. The algorithm acts as a silent assistant of sellers, surfacing timely, eye-catching posts based on previous behavior, allowing buyers to find relevant items without active searching. Sellers create posts with visuals, captions, and timing to maximize reach, and the algorithm uses these signals to target interested buyers. P2 explained this way,

\begin{quote}
\textit{``Since I spend so much time on Facebook, shopping from there is much easier for me. Sometimes in my mind, if I think about one product, that product somehow comes into my newsfeed also, so it is really easier for me to do shopping on Facebook.'' \textbf{(P2)}}
\end{quote}


\textbf{b) Algorithm-Driven Comparison as Buyer Empowerment} Another clear advantage of the algorithm as a seller team member is that it eases comparing products across multiple pages. While the algorithm primarily boosts posts for sellers, buyers benefit by interacting with one or two product posts, which triggers the algorithm to populate their feed with similar items from other sellers. Instead of searching across multiple pages one by one, the algorithm helps them explore a wider range of sellers and price points automatically, making product discovery and decision-making more convenient. P4 described this way,

\begin{quote}
\textit{``I have seen that when I like a product, Facebook shows me similar products from other pages too. I actually use this trick during my shopping because I can visit 3 or 4 pages, check their products, and ask for the price. Then I buy from the one with the lowest price or the product I feel much better. I feel this is actually better than randomly searching for things or buying from only one page." \textbf{(P4)}}
\end{quote}

This interaction shows how buyers can lean on the informal algorithmic boost to create a personalized flow of options tailored to their interests. In this way, the strength of the algorithm as a seller's teammate ends up generating unique advantages for buyers by creating a more dynamic, choice-rich shopping experience that mimics the flexibility of a physical market.

\subsubsection{Shortcomings of Algorithm: Buyer Perspective}
Algorithms also have some clear limitations, including a lack of bargaining flexibility, privacy concerns, and the risk of misleading product recommendations. In this subsection, we explore some of the common challenges that buyers face when relying on algorithms in informal e-markets.

\textbf{a) Lack of Bargaining Flexibility in Automated Chatbots} One of the most commonly mentioned shortcomings of the algorithm from the buyer's perspective is its inability to support bargaining. While sellers use automated chatbots for instant responses, these systems typically provide only predefined answers, like product details or fixed prices. This may seem helpful at first, but when a buyer wants to negotiate, the interaction breaks down. The algorithm lacks the ability to handle the nuances of price negotiation, emotional tone, or contextual decision-making that a human seller could manage. Buyers noted that while chatbots delay significantly once the conversation moves beyond scripted responses, this often leads them to switch to another seller. P4 explained how this often leads him to abandon the purchase.

\begin{quote}
\textit{``When I message a page asking for the product name, I get an automatic reply right away. But when I try to negotiate the price, there is often no response. Sometimes I get a reply after 3 or 4 hours, and by then, I have lost interest and gone to other sellers who reply faster, and I can quickly buy the item. " \textbf{(P4)}}
\end{quote}

In this context, the algorithmic assistant that benefits the seller by handling routine messages ends up being a limitation for the buyer, who expects real-time interaction and bargaining flexibility. Unlike a human staff member who can assess the buyer's tone, willingness to pay, or urgency, the chatbots lack both empathy and improvisation. As a result, a simple interaction can quickly turn into a frustrating and impersonal experience. This reflects a critical mismatch between platform automation and local buyer expectations in informal e-markets.

\textbf{b) Algorithmic Exposure Through Likes and Comments} A major shortcoming of the algorithm in informal e-markets is its inability to protect their privacy. While the algorithm boosts seller visibility by amplifying every like, comment, and interaction, this also makes buyers' activity public. In real-world shopping, choices are private, but online, a simple like or comment can be broadcast to friends' feeds with messages like "Your friend liked this post" or "Your friend commented here." The algorithm amplifies everything indiscriminately, revealing buyers' interests without their control. P20 reflected on how awkward this visibility can become.

\begin{quote}
\textit{"Everyone can see that I showed interest in a product. That makes me feel embarrassed, as I do not want people to know what I am thinking of buying. It just feels too awkward, so I avoid liking or commenting on such posts." \textbf{(P20)}}
\end{quote}

Buyers also observed how the algorithm uses these visible interactions to shape not just their own feed but the feeds of people around them. As friends engage with certain sellers or products, those items start appearing for others as well. This behavior mimics word-of-mouth marketing but without consent. P17 noted how quickly his feed shifts based on what friends engage with:

\begin{quote}
\textit{"I have noticed that if someone likes or comments on a page like that, similar posts start showing up on my feed more often. Sometimes it even says that one of my friends commented on the post. I do not know exactly why this happens, but whenever a friend comments, I start seeing more of those posts." \textbf{(P17)}}
\end{quote}

Our participants noted that this lack of discretion can possibly discourage buyers from engaging with the posts they are interested in, as it undermines privacy and the sense of control over their online activity.

\subsection{Delivery Personnel's Perspectives}
Delivery personnel play a critical role in the informal e-market ecosystem, ensuring that products reach their final destination. They are responsible for picking up orders from centralized hubs and delivering them to scattered customer locations across the city. Unlike traditional logistics systems, these delivery workers operate within a fast-paced, algorithm-driven environment where their daily routines and earnings are heavily influenced by platform algorithms. Despite being essential team members in this digital marketplace, they often face significant algorithmic challenges that disrupt their workflows and reduce efficiency.

A major challenge faced by delivery personnel is related to order assignment and route optimization. Ideally, these algorithms should group deliveries based on geographic proximity to minimize travel time and fuel costs. However, in practice, this system is inconsistent. For example, the platform \textit{Pathao Parcel}, aims to optimize deliveries by assigning orders from the same area to a single delivery worker. However, this approach does not always work as intended. As P35 explained, the algorithm sometimes assigns orders from widely scattered areas, leading to challenging delivery routes and higher operational costs:

\begin{quote}
\textit{“In a day, I typically get 20 to 30 orders. Ideally, all my orders should be in similar areas. When Pathao assigns the orders, it usually groups them by location. However, I often receive random assignments, like 15 orders in one area and another three from different parts of the city. I cannot cancel these orders, as my rating would drop, reducing my future order volume. I really do not know what criteria they use for the assignment, but it causes significant challenges.” \textbf{(P35)}}
\end{quote}

This disconnect between the algorithm's intended efficiency and the practical realities of local delivery work creates substantial barriers for frontline workers, impacting their earnings, physical well-being, and overall job satisfaction.

Another critical challenge is related to delivery confirmation and payment processing. Once a delivery is completed, the driver must update the status in the platform's system to ensure they receive payment. However, our participants noted that this process is not always seamless. The delivery status can remain 'pending' even after the product is successfully handed over to the customer, requiring the delivery worker to take additional steps to verify, as P36 puts it, 

\begin{quote}
\textit{“After the delivery, when I update the status as 'delivered,' it sometimes still shows as 'pending.' If I make 30 deliveries in a day, at least 5 or 6 might still show as pending. To resolve this, I have to visit the hub every evening, provide proof, and manually update the status on their system. Otherwise, I might not get paid for these deliveries, and the system might flag me as if I have stolen those products.” \textbf{(P36)}}
\end{quote}

These challenges demonstrate that while algorithms are essential for scaling and optimizing informal e-markets, they can also create significant operational burdens for delivery personnel if poorly designed and calibrated. Addressing these issues can improve the efficiency and satisfaction of frontline workers, ensuring smoother, more reliable delivery operations and strengthening the overall resilience of the informal e-market ecosystem.

\subsection{AI-based Support Providing Startups' Perspective}

Now we present insights from interviews with entrepreneurs who develop AI tools for informal e-market sellers in Bangladesh. First, we discuss the types of services that sellers commonly seek from AI tools, focusing on the specific features and support that existing platform algorithms like those on Facebook often lack. Second, we explore the technical and resource challenges these startups face, highlighting the barriers they must overcome to effectively serve this rapidly growing sector.

\subsubsection{Seller Requests and Service Offerings by Startups}
AI startups working with informal e-market sellers frequently receive requests for specialized tools that address gaps in existing platform algorithms. Sellers often seek help with tasks like post scheduling, creating better product visuals, scriptwriting, inventory tracking, and emotional tone detection in customer conversations. This section discusses these key service offerings and the reasons why they are in high demand among sellers.


\textbf{a) Post Scheduling and High Engagement Tips} One of the most common seller requests is for tools that optimize post timing and boost engagement. Most small-scale sellers on platforms like \textit{Facebook} struggle to manage consistent, high-visibility content. P19 noted that many sellers request AI-driven tools that can automatically recommend the best times for posting and suggest content types that are likely to perform well. These types of tools are feasible and already exist in some forms, like social media management platforms that use machine learning to predict peak engagement windows. However, in informal e-markets, the data is often unstructured, and audience behavior can be influenced by local events, cultural norms, and platform-specific trends. Despite these challenges, AI startups see the potential in smart scheduling tools, as they can reduce the manual effort of managing page activity, freeing sellers to focus on customer interactions and order fulfillment.


\textbf{b) Better Product Visuals} Another most common seller request is for better product visuals. Many small-scale sellers in Bangladesh rely on basic smartphone cameras and minimal editing tools, which can make their product photos and videos look unprofessional. This can hurt their chances of attracting buyers in a competitive market where first impressions are crucial. P24 noted that here AI tools could provide support,

\begin{quote}
\textit{"Most sellers in our country do not have access to DSLR cameras or professional video editing software. They usually upload regular photos or basic videos. So, we are thinking about how to provide them with high-quality, generated content, like professional-looking photos, edited videos, or polished product images, to help them reach more people and attract buyers more effectively." \textbf{(P24)}}
\end{quote}

To address this, AI startups are exploring solutions like automated image enhancement, background removal, and smart video editing, helping sellers produce professional-quality visuals without expensive equipment or advanced skills. However, this approach presents technical challenges, as algorithms need to understand different product types and cultural preferences, requiring large, diverse training datasets. Startups also face the practical issue of integrating these tools into platforms like \textit{Facebook}, where many sellers operate but API access is limited. 

\textbf{c) Scriptwriting Based on Audience} For AI startups working in the informal e-market, another promising area of innovation is scriptwriting for social media posts. Effective scripts and captions are critical for grabbing attention, building trust, and engaging the right audience, but many sellers lack the time, skills, or experience to craft optimized text. P24 highlighted this need and explained,

\begin{quote}
\textit{"We are also exploring how to improve script writing based on past data, like training the model to generate better captions or video scripts. The idea is to figure out how captions should be written depending on the type of video, so the content reaches more of the target audience. We are also trying to understand how Facebook handles this internally and how elements like well-written captions, emojis, and keywords can help sellers reach more people." \textbf{(P24)}}
\end{quote}

Startups working on this problem often aim to build models that analyze large volumes of past posts to identify patterns in successful content, including the relationship between post structure, tone, and audience engagement. By training their models on these patterns, they hope to offer sellers personalized recommendations that go beyond generic templates. However, this approach has challenges, as informal sellers often lack standardized data, making it hard for AI models to find consistent patterns. Additionally, understanding how \textit{Facebook}'s algorithm ranks and promotes content is complex. Despite these hurdles, AI-driven scriptwriting tools have the potential to help small sellers craft more impactful messages, improving their reach and customer engagement in the crowded digital marketplace.

\textbf{d) Inventory Tracking for High-Volume Sellers} Efficient inventory management is a critical challenge for informal sellers, particularly those handling a large volume of orders. Many small-scale sellers in informal markets rely on manual methods like handwritten logs, spreadsheets, or even memory to keep track of their stock. This approach often leads to mistakes, such as overselling items that are no longer in stock or failing to restock popular products in time. These errors can damage customer trust and hurt long-term business growth. P15 suggested, 

\begin{quote}
\textit{"Sellers who run small accounts or handle a high volume of deliveries usually have a lot of inventory. Manually keeping track of everything becomes really difficult for them. So, if there is a system that can help track their inventory, like how many products came in or are left. We've received this kind of request especially from high-level sellers." \textbf{(P15)}}
\end{quote}

AI startups are focusing on addressing this challenge by developing more flexible, context-aware inventory tools. They are building systems that can automatically track sales activity from social media or messaging apps, using natural language processing to capture order details from customer conversations. Others are integrating with mobile payment systems or digital ledgers, allowing sellers to sync inventory data directly as transactions occur. 

\textbf{e) Emotional Tone Detection in Chat Automation} Understanding the emotional tone of buyers during conversations is a critical challenge for AI startups serving informal e-markets. Unlike formal e-market platforms, where customer interactions are often standardized and transactional, informal markets rely heavily on personal, trust-building communication. Sellers must often interpret subtle cues to assess whether a buyer is satisfied, confused, or frustrated. This emotional context can significantly impact the success of a sale and customer loyalty. P21 described how their team is exploring ways to make chatbots more emotionally responsive, allowing them to detect and adapt to the tone of a conversation in real-time:

\begin{quote}
\textit{"Along with the automatic chatbot, we are also exploring emotional tone detection. If the system senses that a buyer might be upset or frustrated, it can switch to a problem-solving mode, using a tone that tries to calm the situation and offer help. Detecting emotional tone during conversations is very important." \textbf{(P21)}}
\end{quote}

Implementing this kind of emotionally aware chatbot requires advanced natural language processing (NLP) models capable of understanding context and emotional subtext. This is especially challenging Bengali language, where tone can vary significantly based on word choice, sentence structure, and cultural context. Moreover, building these systems requires large amounts of conversational data to train models effectively, and such data is often difficult to obtain in informal e-markets. 

\subsubsection{Technical and Resource Challenges faced by Startups}
While there is significant demand for AI tools in the informal e-market, startups face numerous technical and resource challenges, such as platform constraints, infrastructure limitations, high costs of building low-resource AI systems, and the need for specialized training in local languages and contexts. This section highlights the technical hurdles that entrepreneurs face in developing scalable, effective AI solutions for informal sellers. 

\textbf{a) Platform Constraints} For AI startups supporting informal sellers, one of the biggest challenges is the dependency on third-party platforms like \textit{Facebook}, which were not designed for commercial use at the scale of informal e-markets. Sellers frequently request features like automated post analysis, engagement tracking, etc. These tools often rely on stable data access and consistent platform rules, which are difficult to guarantee on platforms like \textit{Facebook}. Informal e-market platforms like \textit{Facebook}, \textit{Instagram} frequently update their algorithms, data policies, and API structures without prior notice. These sudden changes can disrupt the functionality of custom AI tools, forcing startups to constantly adapt their systems just to keep them operational. P15 explained how these unpredictable shifts can break the trust of sellers.

\begin{quote}
\textit{"We also have to consider the platform we are working on. Platform-related constraints are a big concern, as Facebook can easily change its policies, affecting our tools, data access, and everything we've built. Its system is not really optimized for small sellers, so when they reduce data access or introduce new rules, it often forces us to change our entire approach, making it hard to build a stable solution." \textbf{(P15)}}
\end{quote}

Participants told us that features like real-time inventory updates, post analytics are tightly controlled by \textit{Facebook}’s API, which can restrict or even block access without warning. This creates a major technical challenge, as even a minor change in \textit{Facebook}’s policies can render an entire AI system obsolete overnight. If an algorithm is trained on data that suddenly becomes unavailable or behaves differently due to platform updates, it can lose its accuracy and effectiveness. This adds to the risk for startups, who must constantly retrain their models or redesign their tools to keep up with platforms.

\textbf{b) Infrastructure Limitations} Many AI startup teams also face serious infrastructural barriers while trying to reach informal sellers across Bangladesh. For AI services to operate effectively, consistent internet access is often assumed. However, in many parts of Bangladesh, especially rural and semi-urban regions, sellers experience frequent signal drops, slow loading times, and power cuts. These conditions limit the usability of any tool that depends on continuous background processing or large data transmission. This reality forces developers to rethink their assumptions and adjust their technical designs to suit lower-resource environments. P19 explained this infrastructural challenge,

\begin{quote}
\textit{"There are also infrastructural limitations in our country. Internet connectivity, especially in rural areas, is not very good. For AI-based services to run smoothly, we need real-time data and stable access, which is hard to guarantee in these areas. In cities like Dhaka, this might work, but in more remote regions, it is quite challenging." \textbf{(P19)}}
\end{quote}

These conditions make it difficult for startups to ensure equal service quality across different regions. Some participants' teams are now considering offline-first designs, Bangla-compatible lightweight models, or tools that can work partially without internet. 

\textbf{c) Not Enough Knowledge of Low-Resource AI} Developing AI solutions for informal e-markets in Bangladesh requires a specialized skill set that differs from mainstream AI training. Most globally available resources focus on large-scale, high-resource models for well-funded companies in high-income countries, where power and bandwidth are not major constraints. In contrast, building AI for local sellers means working within low-resource environments, where bandwidth is limited, data is scarce, and language needs are highly localized. AI startup teams in this context often struggle to find the right training and technical resources. While many young developers are technically skilled, their training rarely covers the challenges of building lightweight, offline-capable models for languages like Bangla, including model compression, memory optimization, and power efficiency. Participant P15 emphasized this disconnect,

\begin{quote}
\textit{"Most of the AI training resources we see are actually based on big models, English language, or Western technology. What we really need is training focused on small models, offline inference, or Bangla NLP. But we do not have enough knowledge or resources in these areas, and that makes it quite a challenge to work on these low-resource AI solutions." \textbf{(P15)}}
\end{quote}

There is a growing need for locally relevant educational resources, mentorship programs, and community support networks that can bridge this gap, helping developers move beyond experimental prototypes to scalable, impactful solutions.

\textbf{d) Limited Vendor Capacity to Afford Tech Resources} Developing AI solutions for informal sellers in Bangladesh has financial challenges. Many startups face high costs for cloud computing, server maintenance, data storage, and continuous model training. Unlike large tech companies with substantial resources, these smaller vendors operate on tight budgets, making it hard to invest in advanced AI tools or premium infrastructure. Training a high-quality language model for Bangla or optimizing algorithms for real-time response requires expensive GPUs, scalable cloud services, and significant data processing power. These costs add up quickly, forcing startups to choose between innovation and affordability. Without the financial backing for long-term R\&D, many teams struggle to keep their systems up to date or expand their features. Participant P21 shared how these financial pressures limit their ability to scale and experiment.

\begin{quote}
\textit{"One of our biggest challenges is the cost of running these systems. Cloud servers, model training, and data storage add up quickly. We do not have the budget to run complex models all the time or store massive amounts of data, which is a real barrier for small startups like us trying to build the AI tools that sellers actually want." \textbf{(P21)}}
\end{quote}

This financial constraint means that even when technical skills are available, startups often lack the resources to deploy, maintain, and improve their systems at scale. As a result, they are forced to rely on simpler, less accurate models or cut back on features to keep costs manageable. This makes it harder for them to offer effective AI support to informal sellers, leading to slower growth.

\subsection{Investors' And Market Supporter's Perspectives}
We discussed Bangladeshi informal e-market's growth with two investors who have been in the tech business for more than 25 years, own several tech farms themselves, and invest in new startups frequently. We noted that they have met more than ten such startups and reviewed their proposals. However, they said they could not fund most of them because they found significant vision mismatching, as one of them explained, this way,

\begin{quote}
\textit{"Most proposals come with a too ambitious range of coverage. For example, one proposal I recall came to me a few weeks back that wanted to build bots to automate responses to customer queries promptly, particularly when the business page owner is busy. That would require taking the customer to a separate prompt, and he said this startup will target the population across the globe. Now, one size does not fit all, but it was difficult to convince them. They kept mentioning the US and UK market, however, he did not understand our local market values to begin with." \textbf{(P37)}}
\end{quote}

We draw on the other investor participant's interview, where he explained how most local AI-startups' faulty understanding of informal e-market is unhelpful for the growth of the industry. He also noted that most Bangladeshi e-market customers with money are middle-aged people who trust more in in-person calls over text messages, unlike younger generations, and hence, strategizing AI to communicate with the customer directly might not be the best idea, as he explained.

\begin{quote}
\textit{"The proposals come to me and they kept mentioning the US and UK market and how they have those auto-bots and chat-based services. However, they did not understand our local market values to begin with, where people prefer human staff to take their call and answer their queries. AI can do something else. Even my own business's promotion line is: Our staff pick the customer's complaint calls." \textbf{(P33)}}
\end{quote}

Both the investors shared similar other stories of their evaluation of proposals, and they noted that such a mismatch in values in most local AI-startups' goals and actual local e-market and AI-startups' lack of understanding of the market are rooted in the country's education and tech-training. They further explained that the people in those AI startups are educated and trained in Engineering schools, which still teach a highly Western-centric curriculum. Additionally, the training materials they go through for learning privately are also created and focused on a Western setting, which does not contain the local market's focus. The investors blamed the country's tech education and the overall tech industry's faulty growth plan at the government level for this. 

On the other hand, the female admin who runs a Facebook group of more than 3000 entrepreneurs of informal e-market was pessimistic about the current role of AI and different algorithms in people's businesses. She told us that Facebook's content moderation often gives her member entrepreneurs a hard time, as they reported to her. We quote her, 

\begin{quote}
\textit{"Many in my group are women selling embroidered and hand-painted cloths, or making cakes with custom designs. One of the most common troubles is with the photos they post for promotion. Facebook's content moderation frequently and randomly picks on their images and accuses them of having offensive signs or something absurd. And keep it invisible for an hour until the page owner is done negotiating with the Facebook content moderation policies. Sometimes it takes several days, and in worst cases, they restrict the pages' activities, retrieving which would take multiple days. This is similar to locking down a physical shop for days, imagine the harm to their business caused just for random algorithmic activities." \textbf{(P34)}}
\end{quote}

We further extended the discussion with her on how frequently she noticed her member e-market entrepreneurs used different forms of AI or benefited from the algorithms of different ICTs they used for business. She mentioned that most of her members are homemakers-women-turned-entrepreneurs with low tech skills, and they were still learning from peers in the group how to run the pages and play with the features. She mentioned that less than 10 percent of her members use custom bots and actually know how to play with the algorithms of ICTs in their business. 

%% file: 6.learnsales.tex
\subsection{Note for Algorithms: What Yet to be Learned}
In this subsection, we compare the concurrent algorithmic practices in Bangladeshi informal e-market to renowned market and sales scholarship on key traits that distinguish effective salespeople \cite{hogevold2021b, singh2017thought, vieira2021sales, amor2019skills, Berry1995, 2-goodsale}. Below, we map out which aspects current algorithms in Bangladeshi informal e-markets are lacking and what is needed for a better informal e-market ecosystem. 

\textbf{Personalization and Customer Understanding.} Relationship Marketing Theory emphasizes long-term bonds built through personalized understanding of customers' needs, preferences, and contexts \cite{Berry1995}. Sellers routinely use this approach to signal attentiveness and value. Algorithms are yet to learn to support this by using predictive engagement triggers, lead scoring, and personalized follow-ups based on customer behavior and preferences.

\textbf{Building Trust and Managing Relationships.} According to Commitment-Trust Theory \cite{morgan1994commitment}, trust is foundational to durable commerce. Sellers build this through transparency, responsiveness, and empathy. Algorithms need to mirror these trust-building strategies and detect dissatisfaction, automated feedback highlighting, and loyalty tracking to identify and retain valued customers.

\textbf{Proactive Engagement and Follow-Ups.} High-performing sellers do not wait they initiate conversations, follow up consistently, and nurture leads \cite{mallin2016developing}. Algorithms are yet to learn to replicate this through automated outreach, time-sensitive engagement prompts, and behavioral analytics to prioritize high-potential buyers \cite{mashaabi2022natural}.

\textbf{Adaptive Selling and Flexibility.} Adaptive selling involves adjusting strategies in real time based on customer mood, context, or feedback \cite{hogevold2021b, singh2017thought}. This is especially critical in informal markets where cultural norms and buyer expectations are fluid. Algorithms are yet to learn to emulate this adaptability using real-time sentiment analysis, cultural inference models, and continuous learning to refine responses dynamically.

\textbf{Selling-Related Knowledge.} Expert sellers leverage product knowledge, market awareness, and customer insights to position their offerings effectively. Algorithms are yet to learn to enhance this capability by integrating product databases, purchase history, and market trends to generate timely, informed responses \cite{hogevold2021b}.

\textbf{Self-Leadership and Motivation.} Self-leadership traits such as goal-setting, self-motivation, and discipline are vital for informal sellers operating in high-pressure environments. Algorithms are yet to learn to support this through performance dashboards, behavioral nudges, and reminders that keep sellers organized and focused \cite{singh2017thought}.



%% file: 7.discussion.tex
\section{Discussion}
This paper draws on our interview study and presents how algorithms in ICTs associated with informal e-market influence the industry from sellers', buyers' and other stakeholders perspectives. We have presented advantages, challenges, and postcolonial tensions in the space. Based on our analysis, we first present design implication and broader implications and then propose the \textit{DAIEM} to support docoloniality and informality in this space. 

\subsection{Design Implication}
Below we present and detail some design implication and possible immediate solutions to some challenges of the informal e-market in Bangladesh.  

\subsubsection{Enhancing Chatbot Interactions for Personalized Communication}
In Bangladesh's informal e-market, algorithms and chatbots serve as surrogate team members, streamlining routine communication. Yet our findings show that while efficient in initial exchanges, chatbots struggle with relational tasks like bargaining and product verification that are core practices in trust-building and value negotiation. Unable to adjust prices based on buyer tone, loyalty, or context, chatbots' rigid logic often leads to buyer frustration and lost sales. Similarly, sellers noted that buyers frequently request personalized videos to verify product quality, completing which labor is situated, improvisational, and beyond current algorithmic capacity. We propose designing context-adaptive chatbots that draw from sentiment analysis, interaction history, and categorized content repositories \cite{raju2018contextual, lee2024design}. Advances in NLP, such as Adaptive Contextual Attention Mechanisms (ACAM), offer pathways for more responsive, situated dialogue \cite{melo2023adaptive}. Grounding these systems in vernacular labor practices reframes chatbot design as a site of augmentation—not automation—of informal sellers’ affective, material, and cultural expertise.

\subsubsection{Improving Product Discovery and Search Efficiency}
In informal e-markets, product discovery is hindered by the absence of structured catalogs and standardized listings. Informal sellers and buyers use culturally embedded, context-rich language that resists algorithmic parsing. As a result, buyer queries like ``Eid collection" or ``budget-friendly fashion" often yield irrelevant results, reducing satisfaction and transactional success \cite{zuo2022context}. To address this, we propose a context-adaptive search design that integrates cultural semantics, buyer intent, and local seasonal trends. Algorithms could be trained on regional slang, hashtag use, and vernacular product descriptors to better interpret nuanced queries. Context-adaptive attention mechanisms, for instance, could distinguish between semantically similar but culturally distinct terms. Community-driven signals like trending tags or local shopping behaviors can further calibrate results, ensuring relevance without imposing rigid taxonomies. This approach would reposition search not as a filtering tool but as a culturally situated interface that honors the informal, relational dynamics of postcolonial commerce.

\subsubsection{Strengthening Brand Visibility and Community Engagement}
In informal e-markets, maintaining brand visibility requires constant manual effort across group promotions, community forums, and comment moderation. Sellers track active groups, post frequently, and defend their reputations from harmful comments, the type of tasks that are emotionally taxing and time-intensive. We found the need for algorithmic support that aligns with vernacular promotion practices and brand protection. Algorithms could recommend optimal groups, automate post scheduling, and identify strategic outreach windows, reducing sellers’ cognitive load. On the defensive front, sentiment-aware moderation tools could filter malicious comments, distinguish sincere feedback, and elevate buyer inquiries for safeguarding trust while reducing emotional labor. Drawing from adaptive machine learning models, such systems could learn from past interactions to tag returning customers or suggest personalized follow-ups. Thus, these tools would augment sellers' grassroots promotional work and support relational commerce in a context-aware manner.

\subsubsection{Real-Time Price Tracking and Personalized Notifications}
In informal e-markets, timely communication about price and stock changes is vital but often handled manually, placing cognitive strain on sellers and leading to missed opportunities. Unlike formal platforms, informal systems lack automated tools to track buyer interest or notify customers about restocks or price drops. We propose context-aware algorithms that monitor inventory, flag changes, and deliver targeted updates to interested buyers, drawing on interaction histories and purchase intent \cite{raju2018contextual, lee2024design}. This would reduce manual follow-up and foster repeat sales. Additionally, our findings reveal delivery workers also face algorithmic inefficiencies: platforms like Pathao assign orders across scattered zones, ignoring geographic logic. Refusing these tasks harms workers' ratings. We advocate for delivery algorithms that allow task preview, flagging, and route-based fairness. By automating communication and redesigning assignment systems, algorithms can shift from rigid enforcers to responsive collaborators in sustaining the informal e-market ecosystem.

\subsection{Broader Implications}
Our study also contribute the domain in theory, critical thinking, and design practice end. Below we discuss the broader implications of this work. 

\subsubsection{Trust and Algorithm for Informal E-market}
Our research joins the growing body of trust scholarship in HCI, social computing, and AI. Most of today's dominant literature focuses on how users form, negotiate, and act upon trust in sociotechnical systems \cite{singh2014norms,harding2015hci, bickmore2001relational,riegelsberger2005mechanics}. Trust in intelligent systems has been widely studied in HCI and AI, with researchers examining how features like transparency, explainability, and reliability influence users' willingness to rely on or question algorithmic decisions \cite{ferrario2022explainability, von2021transparency, schmidt2020transparency,kaur2022trustworthy}. Our research extends this chain of literature by bringing in questions of how we can foster interpersonal trust when algorithm has a significant role in the interaction happening in a transactional ecosystem. While research at the intersection of algorithm, trust, and market grown toward making systems and infrastructure more trustworthy, our research shows that market in the Global South need ways to make interpersonal relations and interactions more trustworthy, in addition. We argue that designers and practitioners working in this space need to widen their knowledge beyond rationality in HCI- and AI-design and work in integrating more hope- and care-based sensitivities, particularly in low-resource contexts like informal e-markets in the Global South \cite{ratto2023reopening, croon2022thinking}.  

\subsubsection{Contextual Learning}
Our research also joins the growing research on contextual learning of algorithms and AI has grown significantly in HCI, CSCW, and AI ethics. This domain has investigated how people come to understand and adapt to algorithmic systems through situated everyday experiences and observations \cite{eslami2016first, devito2017algorithms, qi2011exploring}. In our work, we found that people's decision-making in informal e-market activities is hugely influenced and shaped by their situated knowledge, hunches, and gut feelings. This aligns with existing research's findings that suggested that people in the Global South market are influenced by their local rumors, customs, and contextual knowledge that is rooted in their culture and myths \cite{rohanifar2022role, sultana2021dissemination, chandra2019rumors, sultana2021chasing}. This set of sentiments is highly complex to capture through sensors and create models for algorithms. We argue that this is a shortcoming in current AI and algorithmic design that still fails to serve a huge number of people who rely more on faith, values, and customs in transactions and other similar decision-making.  

\subsubsection{Postcolonial Vision Mapping for the Market}
Our work also joins the postcolonial computing movement within HCI and ICTD \cite{124irani2010postcolonial, 125merritt2011postcolonial,117ahmed2015residual}. Building on the concept of `otherness' \cite{127said1995w}, postcolonial computing demonstrates how mainstream computing knowledge is often ignorant of local knowledge, and that creates a space for postcolonial struggle through marginalization, resistance, disapproval, and failure \cite{128philip2012postcolonial,129ahmed2017digital, sultana2019witchcraft, sultana2021chasing, sultana2021dissemination}. These studies connect themselves to a rich body of work in history and social science depicting how colonized knowledge has been historically marginalized and suppressed, and colonized bodies, materialities, and interaction are often neglected, examined, marked, exploited, or deprived \cite{132arnold1993colonizing, 131spivak2003subaltern, 73spivak1988can}. This paper joins this body of work by bringing insights into the AI-related market's growth in the Global South. Our investigation with the entrepreneurs who build AI bots and other support systems for small-scale informal e-markets and their investors in Bangladesh showed that the way entrepreneurs are trained and motivated is very west-centric, which was a mismatch with how the investors envisioned the possible growth trajectory of the market. For example, the entrepreneurs were highly interested in automating the processes as much as possible, while investors, being in the market for more than 25 years, informed us that their glory was about having complaint calls picked up by human staff at their office. Additionally, the investors informed us that they noticed entrepreneurs' visions in designing affordances for the AI and other related business tools had a significant Western influence. The investors argued that many of their planned products would fail in the market because they do not take people's values and local infrastructure into account. This is particularly because the young entrepreneurs are trained at their local formal educational institutions (schools and colleges) and from informal learning sources (online educational and training materials), all of which are still highly influenced by Western knowledge that ignores the local knowledge by `othering' them. To have a sustainable informal e-market ecosystem in Bangladesh and similar other settings, we must think beyond just algorithm design and focus on decolonizing the entities and stakeholder actors in the infrastructure.

\subsection{Toward \textit{`Decolonial Algorithm for Informal E-market' (DAIEM)}}
Our findings and further analysis led us to develop a framework for designing and evaluating algorithms in ICTs associated with informal e-markets. We name it the \textit{``decolonial algorithm for informal e-market'' (DAIEM)}. This framework would work as both a critical guideline for designing informal e-market-oriented infrastructure and an analytical tool for evaluating such ecosystems. We believe the \textit{DAIEM} framework will be useful to HCI, algorithm, and AI designers and practitioners working to sustain informal e-market in Bangladesh and similar other settings, and can also be adaptive and expandable beyond the algorithm in the market context to other contexts such as algorithms for collaboration in global development, informal care, and other sectors that run on imbalances of power and low-resources. \textit{DAIEM} framework's decoloniality and informality sentiments are supported through six components, including autonomy and agency; resistance; locality, culture, and history; rationality; materiality; and advocacy. We discuss them below:

\subsubsection{Autonomy and Agency}
\begin{wrapfigure}{r}{0.5\textwidth}
\vspace{-10pt}
\includegraphics[width=0.9\linewidth]{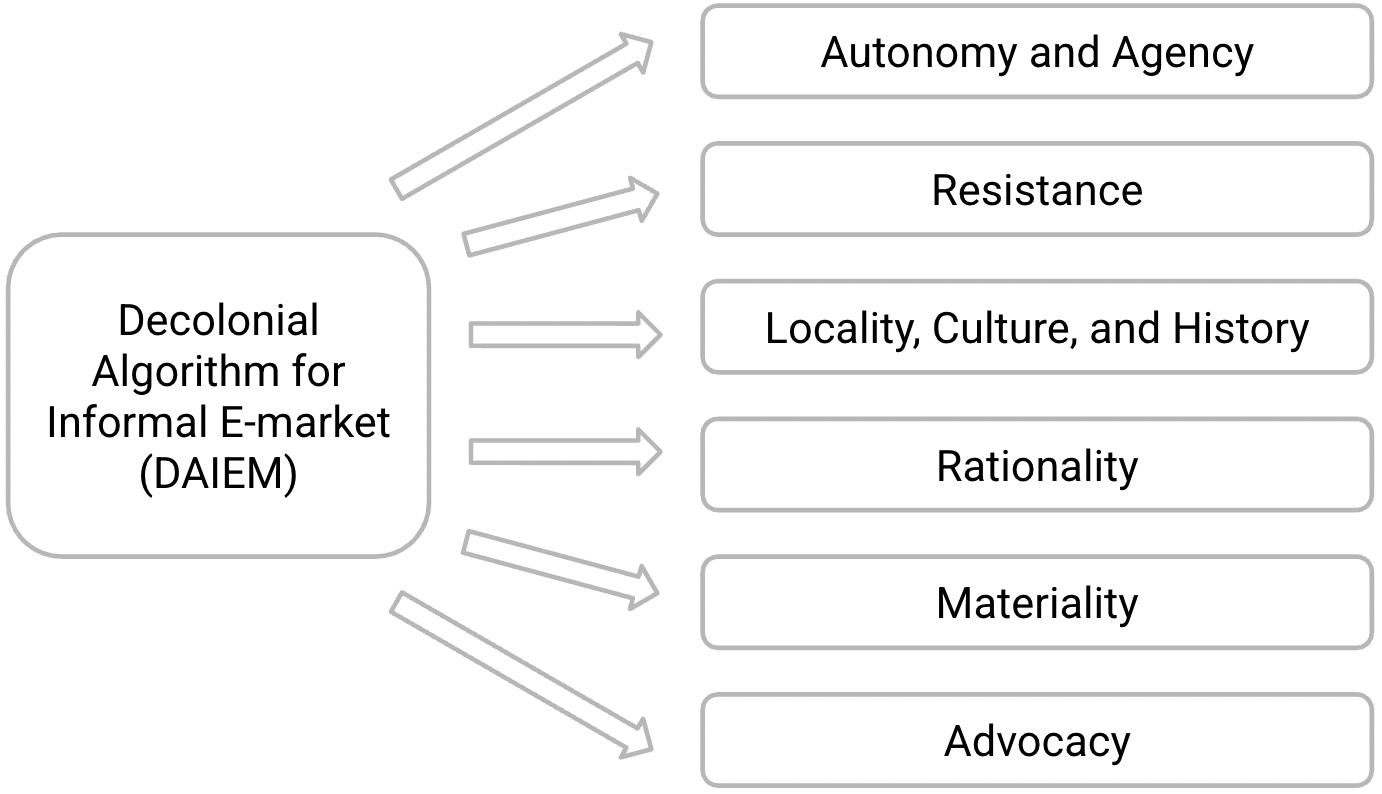} 
\caption{The \textit{Decolonial Algorithm for Informal E-market (DAIEM)} framework consisting of six dimensions: autonomy and agency; resistance; locality, culture, and history; rationality; materiality; and advocacy.}
\label{fig:wrapfig}
\vspace{-15pt}
\end{wrapfigure}
Designing algorithms for informal e-markets in postcolonial contexts requires a shift from universalist optimization logics toward infrastructural arrangements that recognize autonomy as situated and contested. Rather than enforcing platform-native metrics of efficiency, credibility, or engagement, algorithmic systems should be responsive to relational trust signals, such as conversational reciprocity, temporal rhythms of exchange, and forms of visibility. Informality must be treated as a dynamic mode of infrastructuring, and algorithmic design should adapt to different modes based on users' contextual needs. A decolonial algorithmic design approach thus demands legibility, reversibility, and user participation to foster interpretability and epistemic agency.

\subsubsection{Resistance} 
Designing algorithms for decoloniality and informality needs to recognize resistance to mainstream market infrastructures not as friction but as a form of design critique enacted through everyday workaround and refusal. Rather than correcting deviation from platform norms, algorithms should accommodate practices like price withholding, conversational bargaining, and intentional opacity as meaningful expressions of market logic. Designing with resistance means foregrounding spaces for indeterminacy, where users can negotiate visibility, delay engagement, or subvert ranking mechanisms without penalization. Such systems reposition algorithms not as enforcement tools of platform governance, but as malleable infrastructures co-shaped by postcolonial users' situated knowledge, informal labor, and collective refusals.

\subsubsection{Locality, Culture, and History}
Algorithms in ICTs for informality and decoloniality need to understand the politics of location, culture, and history of the data to be able to address the concerns associated with human, their interaction with ICTs, and their data. This would require growing contextual and situated knowledge in associated informal practices, which can be obtained by adapting to more epistemologically rich methods. The politics of context is also complicated, and current AI, critical data studies, and HCI scholarship do not know how to valuate postcolonial data as of the techniques that are dominantly west-influenced --- which is another major philosophical shortcoming of the domain.

\subsubsection{Rationality} 
Dominant algorithms in informal e-market ICTs often rely on Western scientific, ethical, and algorithmic rationality approaches that emphasize quantification, risk prediction, and standardized outcomes. However, such frameworks can obscure or delegitimize other forms of knowing and rational decision-making that are deeply embedded in users' cultural, spiritual, emotional, and collective experiences, i.g, \textit{alternative rationalities} \cite{sultana2019witchcraft, sultana2021dissemination, sultana2020parareligious, sultana2021chasing}. Algorithms that ignore these forms of reasoning risk epistemic violence, rendering users' practices invisible or deviant. To design algorithms for decolonial and informal systems, HCI, CSCW, and AI design must expand their epistemic scope, honor and integrate plural rationalities, and create tools in design and practice that support without overwriting how people make sense of their postcolonial experiences.

\subsubsection{Materiality}
Designing decolonial algorithms for informality would also require engaging with the material infrastructures (e.g., devices, languages, currencies, and delivery networks) and understanding how they shape postcolonial users' informal transaction behavior and communication. Rather than abstracting these behaviors into data points, algorithms should be sensitive to the frictions of low-bandwidth messaging, cash-based exchanges, shared phone use, and asynchronous delivery rhythms that define informal economies. Note that informality is a material condition that demands systems capable of operating through partial connectivity, improvisation, and tactile trust signals. Algorithmic design, then, must move beyond disembodied optimization to support grounded, infrastructurally entangled practices that reflect the realities of postcolonial material life.

\subsubsection{Advocacy}
A decolonial approach to algorithm design for informal e-markets needs to center the advocacy practices of postcolonial users and stakeholders who actively shape, contest, and reimagine platform norms from the margins. Rather than treating advocacy as external to system design, algorithms should be responsive to collective demands, such as calls for fair visibility, local dispute resolution, and culturally relevant features emerging through user networks, public discourse, and informal feedback loops. Designing for advocacy means building infrastructures that support negotiation, allow for contestation, and encode accountability not just to system metrics but to the lived realities of sellers, buyers, and community intermediaries. In doing so, algorithms become sites of political co-production, reflecting not only technical logics but also the ongoing struggle for identity, recognition, fairness, and representational power in postcolonial digital economies.



\subsection{Limitations of the Work}
While our findings are grounded in rich qualitative data from diverse stakeholders within Bangladesh's informal e-market, our work has several limitations. First, the study is limited to participants who were already active on digital platforms and may not capture experiences of those excluded due to infrastructural, linguistic, or financial barriers. Second, our interpretations are shaped by the specific sociopolitical and technological context of Bangladesh, which may not generalize to informal e-markets in other postcolonial regions. Additionally, the reliance on interviews limits our ability to observe long-term algorithmic adaptations or invisible labor behind digital maintenance. Finally, while our proposed \textit{DAIEM} framework draws from grounded insights, it remains a conceptual contribution that requires further empirical validation through design and deployment.

\section{Conclusion}
This study contributes to CSCW and HCI by foregrounding the algorithmic experiences of informal e-market actors in postcolonial contexts, centering Bangladesh as a site of situated digital practice. Through interviews with buyers and sellers, we show how participants navigate opaque platform logics, enact relational trust, and reconfigure global tools to meet vernacular needs. Rather than seeking formalization or scalability, participants prioritize autonomy, visibility, and cultural legibility. Our proposed \textit{DAIEM} framework offers a conceptual foundation for rethinking algorithmic systems not as universal solutions but as infrastructures that can be co-shaped by localized practices, resistances, and material constraints. We argue that decolonial algorithm design must account for the cultural, infrastructural, and political dimensions of informality of the market. 